\documentclass[runningheads]{llncs}
\usepackage{graphicx}
\begin{document}
\title{Inter Subject Emotion Recognition Using Spatio-Temporal Features From EEG Signal\thanks{All the authors have equal contributions.}}
%
%
\author{Mohammad Asif\inst{1}\orcidID{0000-0002-9517-6716} \and
Diya Srivastava\inst{1} \and
Aditya Gupta\inst{1} \and Uma Shanker Tiwary\inst{1}\orcidID{0000-0001-7206-9013}}
\authorrunning{M. Asif et al.}
%
\institute{Indian Institute of Information Technology, Allahabad, UP, India 
\email{pse2017001@iiita.ac.in}}
\maketitle All the authors have equal contributions.              
\begin{abstract}
Inter-subject or subject-independent emotion recognition has been a challenging task in affective computing. This work is about an easy-to-implement emotion recognition model that classifies emotions from EEG signals subject independently. It is based on the famous EEGNet architecture, which is used in EEG-related BCIs. We used the 'Dataset on Emotion using Naturalistic Stimuli' (DENS) dataset. The dataset contains the 'Emotional Events'- the precise information of the emotion timings that participants felt. The model is a combination of regular, depthwise and separable convolution layers of CNN to classify the emotions. The model has the capacity to learn the spatial features of the EEG channels and the temporal features of the EEG signals variability with time. The model is evaluated for the valence space ratings. The model achieved an accuracy of 73.04\%.

\keywords{Affective Computing  \and Emotion Recognition \and DENS Dataset \and EEG \and CNN \and Subject Independent.}
\end{abstract}
\section{Introduction}

Affective computing is a rapidly emerging field due to the fact that it has the potential to transform how humans-humans, humans-machines and machines-machines interact with each other. It is an amalgamation of concepts from diverse fields such as psychology, computer science and engineering, neuroscience etc. It primarily focuses on developing technologies that can detect, interpret, and respond to human emotions.  A variety of modern-learning technologies, such as machine learning, deep learning and natural language processing, brain signals etc., are used in affective computing. We have used a deep learning model to do our emotion recognition task from electroencephalogram signals\cite{verma2017affect}.

Emotion recognition has never been an easy task and has gained major attention and prominence in the past few years\cite{kaur2021emotion}. Out of the various techniques adopted, electroencephalography(EEG) has been considered one of the most reliable methods. ‘Dataset on Emotion with Naturalistic Stimuli’ (DENS dataset)\cite{10101783}, indigenously recorded by us, is used for the underlying study. 

Inter-subject or subject-independent EEG-based emotion classification involves the classification model learning and identifying the features that are common across different individuals without relying on information specific to any particular individual. 
Subject-independent EEG-based emotion classification has often been perceived as a much more challenging and complex task, primarily due to the invariability of EEG signals across different subjects. Hence, it draws less accuracy than within-subject or subject-dependent emotion recognition.
There are various methods and deep learning architectures based on CNN, RNN, DNN etc., available for emotion recognition that works well on subject-dependent emotion tasks \cite{siddiqui2023deep}\cite{craik2019deep}\cite{7822545}.

\subsection{Literature Survey and Challenges}
The emotion recognition task for subject-independent data is much more complex than for the subject-dependent data as invariability exists in the brain activity of two individuals in spite of subjecting them to the same stimuli and under similar conditions. The accuracy obtained in subject-independent data-based emotion recognition falls short in comparison to that of subject-dependent. Using the LSTM model on the DEAP dataset , for subject-independent study, an accuracy of 70.31\% on the valence space and 69.53\% on the arousal space is obtained, which is lower by nearly 30+ percent than that in the subject-dependent study\cite{nath2020comparative}. EEG signals are raw signals from the brain and undergo various preprocessing steps before being given as input to the model. Transforms such as wavelet transform\cite{pandey2019subject} and Short Time Fourier Transform\cite{pandey2019emotional} are employed for the purpose of feature extraction from the preprocessed EEG data. A proposed Deep Neural Architecture, which takes data whose feature extraction is done via wavelet transform for subject-independent study, gives an accuracy of 62.5\% for the valence space and 64.25\% for the arousal space. The emotion recognition accuracy also depends on the electrodes and the bands used. Multilayer perceptron gives good accuracy on the DEAP dataset (subject-independent study), and this value differs as we switch electrodes\cite{pandey2019emotional}. For DEAP and SEED datasets, using the CNN-GRU model and using variable EEG bands, an average accuracy of 67.36\% for valence on the subject-independent DEAP dataset and 70.07\% for the same in arousal. For the SEED dataset,  an average accuracy of 87.04\% is obtained\cite{xu2023subject}.

Here, we use an easy-to-implement and compact model for emotion recognition based on CNN. It automatically learns fractures from raw EEG signals without the use of any explicit feature extraction techniques such as continuous wavelets (CWT), Fourier transforms (STFT)\cite{asif2023emotion} etc.

\begin{figure}
\includegraphics[width=\textwidth]{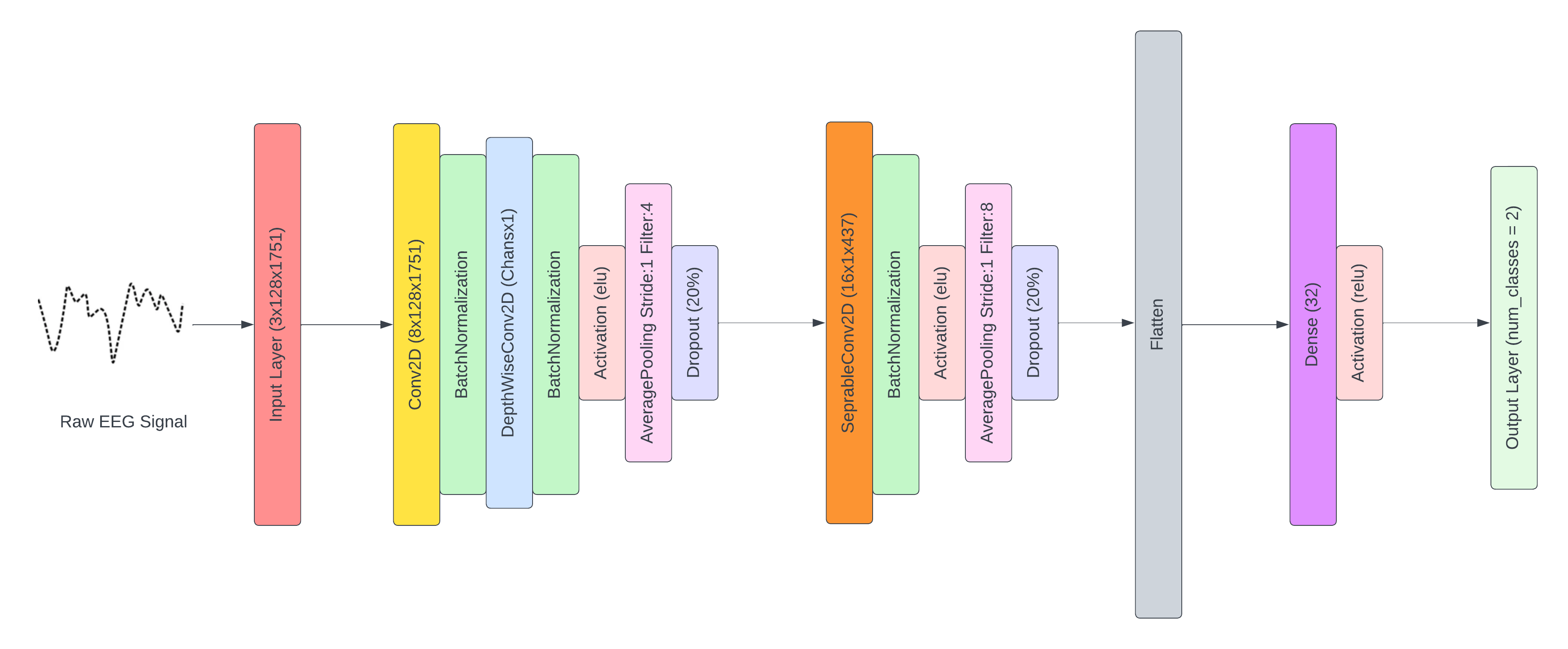}
\caption{Architecture of the Proposed Model.} \label{model}
\end{figure}

\section{Data and Methodology}
\subsection{Dataset Description}
Various EEG datasets are available for emotion \cite{koelstra2011deap}\cite{zheng2015investigating}\cite{7887697}\cite{8554112}, but precise information about when emotion was felt is missing in these datasets. We have used the DENS\cite{10101783} dataset for the classification task. The dataset contains information on when participants felt the emotion. It used 128 channels of high-density EEG recording.

The dataset contains 465 emotional events extracted from 40 participants. The data is pre-processed with a bandpass filter (1-40 Hz) and Independent Component Analysis (ICA) \cite{brainsci12060702}. The data signal length for each emotional event is 7 seconds which has 125 Hz sampling. There are 128 channels in the EEG.
\subsection{Model Specifications}
We have used architecture similar to EEGNet for our subject-independent emotion classification task because of its compactness and generalization ability. EEGNet is a compact CNN for the classification and interpretation of EEG-based Brain-Computer Interfaces. It uses a depthwise convolution layer and separable convolution layer along with various traditionally used layers in its architecture. We tweaked the architecture according to the improvements needed. 

The depthwise convolution gets its name because it primarily works with the width and height of an image and kernel. The number of channels in each image makes up the second dimension, or "depth" dimension. A kernel is simply split into two smaller kernels in a spatial separable convolution. Using this convolution layer, the computational complexity goes down. This layer not only deals with the spatial dimensions but also with the depth dimension, which is actually the number of channels.

The separable convolution layer is a combination of the depthwise convolution layer and pointwise convolution layers. It helps to lower the trainable parameters and also works well for the time variability of the signals.

This model works well with other BCI data; however, it gives less accuracy in the case of emotions. We tweaked the input formatting and also added an extra dense layer to train our model. The overall architecture is mentioned in Fig.\ref{model}, and other parameters are mentioned in Table \ref{parameters}.

\begin{table}
\centering
\caption{Parameter Settings of the Model.}\label{parameters}
\begin{tabular}{|c|c|}
\hline
{\large\bfseries Parameters} &  {\large\bfseries Settings}\\
\hline
{\bfseries Optimizer } &  Adam\\
{\bfseries Loss Function} &  Sparse Categorical Cross-entropy\\
{\bfseries Learning Rate} & 0.01\\
{\bfseries Adjustment} & Early Stopping Criteria: Monitor – ‘val\_loss’: patience = 35 Model Checkpoint – ‘val\_loss’\\
{\bfseries Activation Function} & CNN Layers: ‘ELU’, Dense: ‘ReLU’ and Output: ‘Softmax’\\
{\bfseries Epochs\cite{xu2023subject}} & 200\\
\hline
\end{tabular}
\end{table}

\section{Results}
The model is evaluated on two classes that are based on the valence values- high and low ratings (ratings from 1 to 5 are considered as ‘Low’, and 5 to 9 as ‘High’) received by the participants (The valence rating ranges from 1 to 9). To ensure the subject independency of the experiment, no intersection of the participants was in the train and test data split.

The performance of the suggested model has been found to be better than the EEGNet. The original EEGNet model gives an accuracy of 56.38\%. The F1 score was found to be 69.63\%. On tweaking the model, the accuracy improved to 73.40\% and the F1 score improved to 82.27\%. The results are on par with the existing models.

The comparison of this model with various other models is shown in Table \ref{comparison}. 

\begin{table}
\centering
\caption{Result Comparison With Various Models.}\label{comparison}
\begin{tabular}{|l|l|l|l|}
\hline
{\large\bfseries Method} &  {\large\bfseries Emotion Space}   & {\large\bfseries Dataset} &  {\large\bfseries Accuracy}\\
\hline
{\bfseries LSTM\cite{nath2020comparative} } &  Valence & DEAP & 70.31\%\\
{\bfseries LSTM\cite{nath2020comparative} } &  Arousal & DEAP & 69.53\%\\
{\bfseries DNN\cite{pandey2019subject}} & Valence & DEAP & 62.50\%\\
{\bfseries DNN\cite{pandey2019subject}} & Arousal & DEAP & 64.25\%\\
{\bfseries CNN-GRU\cite{xu2023subject}} & Valence & DEAP & 67.37\%\\
{\bfseries CNN-GRU\cite{xu2023subject}} & Arousal & DEAP & 70.07\%\\
{\bfseries CNN-GRU\cite{xu2023subject}} & Valence & SEED & 87.04\%\\
{\bfseries EEGNet\cite{lawhern2018eegnet}} & Valence & DEAP & 56.38\%\\
{\bfseries Our Model} & Valence & DENS & 73.40\%\\
\hline
\end{tabular}
\end{table}

\section{Conclusion and Future Aspects}
The proposed architecture is light weighted and easy to implement. For the subject-independent emotion recognition problem, good accuracy and F1 score are obtained with considerably less number of learnable parameters which prevents the model from being overtrained. Getting good results is challenging and due to limited data becomes even harder. The model is getting good results as the different layers are able to extract the temporal as well as spatial features from the EEG signal.
Since EEG contains more noise than the data, it becomes challenging for the model that discards irrelevant information from the signal. As precise information is available for the DENS data, it benefits the model to learn emotion-specific EEG activity only, which is essential for a subject-independent or inter-subject study.

The model will be, in future, experimented on the Arousal and Dominance spaces. Model validation with more datasets for emotion analysis also lies in the future prospects. In our previous work for subject-dependent emotion analysis, we found that adding the information of the subjective feelings (Dominance) enhances the accuracy\cite{asif2023emotion}. So in future work, consideration of the dominance ratings may also improve the accuracy of the model.

%
%
%
%

\end{document}